\documentclass[prb,aps,twocolumn,showpacs,showkeys]{revtex4}
\usepackage[english]{babel}
\usepackage{epsfig}
\usepackage{amsmath}
\usepackage{amssymb}
\usepackage{amsfonts}
\usepackage{bm}

\begin{document}

\title{Small-scale phase separation in doped anisotropic antiferromagnets}

\author{M.Yu.~Kagan,}
\affiliation{Kapitza Institute for Physical Problems, Russian
Academy of Sciences, Kosygina str. 2, Moscow, 119334 Russia}
\author{K.I.~Kugel, A.L.~Rakhmanov,}
\affiliation{Institute for Theoretical and Applied
Electrodynamics, Russian Academy of Sciences, Izhorskaya str.
13/19, Moscow, 125412 Russia}
\author{K.S.~Pazhitnykh}
\affiliation{Moscow Engineering Physics Institute (State
University), Kashirskoe shosse 31, Moscow, 115409 Russia}

\begin{abstract}
We analyze the possibility of the nanoscale phase separation
manifesting itself in the formation of ferromagnetic (FM) polarons
(FM droplets) in the general situation of doped anisotropic three-
and two-dimensional antiferromagnets. In these cases, we calculate
the shape of the most energetically favorable droplets. We show
that the binding energy  and the volume of a FM droplet in the
three-dimensional (3D) case depend only upon two universal
parameters $\bar{J} =(J_x + J_y + J_z)S^2$ and $t_{eff}
=(t_xt_yt_z)^{1/3}$, where $\bar{J}$ and $t_{eff}$ are effective
antiferromagnetic (AFM) exchange and hopping integrals,
respectively. In the two-dimensional (2D) case, these parameters
have the form $\bar{J} =(J_x + J_y)S^2$ and $t_{eff}
=(t_xt_y)^{1/2}$. The most favorable shape of a ferromagnetic
droplet corresponds to an ellipse in the 2D case and to an
ellipsoid in the 3D case.

\end{abstract}

\pacs{71.27.+a,
64.75.+g,
75.30.Gw,
75.47.Lx
}

\keywords{electronic phase separation, magnetic semiconductors,
magnetic polaron}

\date{\today}

\maketitle

\section{Introduction}

The problem of electronic phase separation with the formation of
ferromagnetic (FM) or paramagnetic (PM) spin polarons (magnetic
droplets or ferrons) due to the self-trapping of charge carriers
in an antiferromagnetic (AFM) matrix became very popular nowadays,
especially in the studies of high-T$_{\mathrm c}$ superconductors
and the systems with the colossal magnetoresistance (such as
LaMnO$_3$ manganites doped by Ca). For isotropic materials, the
size and shape of FM droplets was evaluated in several papers
beginning from the seminal work of Nagaev~\cite{Nag67}, for more
detail see Ref.~\onlinecite{KaKhMo}. The characteristic size of a
FM droplet turns out to be of the order of 15--20\AA \,and its
optimum shape in isotropic 3D manganites is a spherical one. Later
on, Kagan and Kugel~\cite{KaK} analyzed the case of layered
manganites (like(La,Ca)$_{n+1}$Mn$_n$O$_{3n+1}$) and demonstrated
that the droplets with the lowest energy have the ellipsoidal
shape. The FM droplets of cylindrical shape considered first by
Nagaev~\cite{Nag99} for this class of manganites correspond to a
higher energy. Currently, the phase separation in anisotropic
materials was also addressed in connection with low-dimensional
organic compounds~\cite{low-d} and quasi-one-dimensional magnets
such as BaCoO$_3$~\cite{BaCoO3,BaCoO3a}. Magnetic polarons in
doped one-dimensional (1D) AFM magnetic chains were recently
considered in Refs.~\onlinecite{Ivan1,Fer1D,Ivan2}, where the
possibility of rather long-range magnetic distortions around the
polaron was demonstrated.

Another possibility to have a strongly anisotropic situation
arises when we take into account an interplay between the
microscopic phase separation and charge ordering (stripe
formation), include the Jahn-Teller type of effects (orbital
degrees of freedom), or consider stable crystallographic
distortions. In these cases, the quasi-1D zig-zag or ladder
structures are often observed in the corresponding
systems~\cite{Khoms,Hotta}.

In this paper, we present the calculations concerning the shape
and size of FM droplets in anisotropic two-dimensional (2D) or
three-dimensional (3D) cases when, generally speaking, the
electron hopping integrals $t_x$, $t_y$, and $t_z$ along $x$, $y$,
and $z$ directions, as well as the AFM exchange integrals $J_x$,
$J_y$, and $J_z$ are different. We get that, by analogy with the
situation in layered manganites~\cite{KaK}, the most favorable
shape of a FM droplet is an ellipsoidal one. Moreover, the binding
energy and the effective volume of the droplet are expressed only
in terms of universal averaged parameters $\bar{J} = \left(J_x +
J_y + J_z\right)S^2$ and $t_{eff} = \left(t_xt_yt_z\right)^{1/3}$.
These results are interesting, in particular, in relation to the
neutron scattering experiments giving an indication of the
existence of FM clusters  with different shapes in perovskite and
layered manganites~\cite{HennPRL98,HennPRB00,HennNJPh05}.

Our paper is organized as follows. First, we consider the purely
2D situation and find the most favorable shape of a 2D ferron
comparing the energies of elliptical and rectangular droplets in
the general anisotropic  2D case: $t_x \neq t_y$ and $J_x \neq
J_y$. We find that in two dimensions, the minimal energy
corresponds to the elliptical shape. Then, we include the third
dimension ($J_z$ and $t_z$) and compare the energies of the
cylinder and ellipse in the case when both of them have the
optimum elliptical shape of the 2D cross-sections. We find again
that the minimal energy in the 3D case corresponds to the
ellipsoidal shape of FM droplets. At the end of the paper, we
provide some discussions and conclusions.

\section{The shape of FM droplets in the anisotropic 2D case}

\begin{figure}
\centerline{\includegraphics[width=0.38\textwidth]{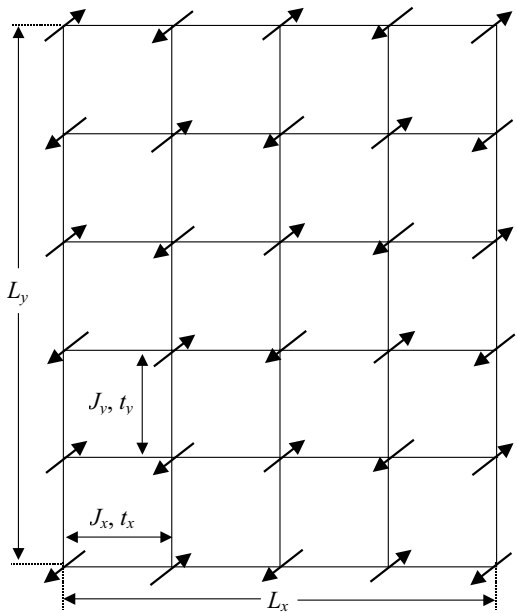}}
\caption{\label{2Dlatt}  2D anisotropic doped antiferromagnet with
the square lattice. $J_x \neq J_y$ are AFM exchange integrals,
$t_x \neq t_y$ are the electron hopping integrals, $L_x$ and $L_y$
define the volume of the rectangular ferron in 2D.}
\end{figure}

Let us first consider the anisotropic 2D case. In this case, there
are two different electron hopping integrals $t_x \neq t_y$ and
two  different constants of the AFM exchange interaction $J_x \neq
J_y$, see Fig.~\ref{2Dlatt}. To some extent, this case has a lot
of similarities with two-leg ladder systems rather popular
nowadays (see, for example, Ref.~\onlinecite{ladder} and
references therein).

Throughout this paper, we consider the Kondo-lattice model with
the Hamiltonian
\begin{equation}\label{Kondo-latt}
\hat{H} = J_H \sum_{i}{{\bf S}_i{\bf \sigma}_i} +
\sum_{\langle{ij}\rangle_{\alpha}}{J_{\alpha}{\bf S}_i{\bf S}_j} +
\sum_{\langle{ij}\rangle_{\alpha}}{t_{\alpha}c_i^{\dag}c_j},
\end{equation}
where, $c_i^{\dag}$ and $c_i$ are electron creation and
annihilation operators at site $i$, $\alpha = \{ x,y \}$ for a
square lattice in 2D, $\langle{ij}\rangle_{\alpha}$ denote the
neighboring sites in the lattice along the $\alpha$ direction,
${\bf \sigma}_i = c_i^{\dag}{\bf \sigma}c_i$ is the spin of a
conduction electron (${\bf \sigma}$ is the Pauli matrix), ${\bf
S}_i$ is a local spin, $J_{\alpha}$ are AFM exchange integrals,
$t_{\alpha}$ are the hopping integrals for conduction electrons,
and parameter $J_H$ corresponds to the Hund's rule coupling
between a local spin ${\bf S}$ and a spin of a conduction
electron.

We work in the double-exchange limit, which implies that $J_H \gg
\{ t_x, t_y \} \gg \{ J_x, J_y \}$. In this case, the ground state
of the system is unstable toward the nanoscale phase
separation~\cite{Nag67,KaKhMo,KaK} with the formation of FM
polarons inside the AFM matrix. Let us now evaluate the total
energy of the phase-separated state for different shapes of
ferrons possible in the 2D case.

\subsection{A rectangular ferron}

Let us first consider a rectangular FM droplet (ferron) located
at the square lattice with the intersite distance $a$. Its
characteristic sizes along $x$ and $y$ axes are $L_x$ and $L_y$,
respectively. The dimensionless volume $\Omega$ of such a ferron
can be defined as $\Omega =L_xL_y/a^2$. The kinetic energy of
charge carriers (electrons or holes) within the FM droplet is
\begin{equation}\label{E_kin}
E_{kin} = -2t_xn - 2t_yn + \varepsilon_0n,
\end{equation}
where $n$ is the concentration of charge carriers and
$\varepsilon_0$ is a binding energy corresponding to the first
(the lowest) level in the rectangular potential well. The latter
can be found by solving the corresponding Schr\"{o}dinger equation
(see Ref.\onlinecite{KaK})
\begin{equation}\label{SchrEq}
\hat{H}_{kin}\Psi(x,y) = \varepsilon_0\Psi(x,y),
\end{equation}
where
\begin{equation}\label{H_kin}
\hat{H}_{kin} = -a^2\left( t_x\frac{\partial^2}{\partial x^2}
+t_y\frac{\partial^2}{\partial y^2}\right).
\end{equation}

For the case of a well-defined ferron (without an extended tail of
magnetic distortions, which we will consider throughout the
present paper, the corresponding boundary conditions have the form

\begin{equation}\label{bound_cond}
\Psi(x =L_x,y) = \Psi(x,y=L_y)= 0.
\end{equation}
Hence,

\begin{equation}\label{Psi(x,y)}
\Psi(x,y) = \sin\frac{\pi x}{L_x}\sin\frac{\pi y}{L_y}
\end{equation}
and
\begin{equation}\label{eps_0}
\varepsilon_0 = t_x\left(\frac{\pi a}{L_x}\right)^2 +
t_y\left(\frac{\pi a}{L_y}\right)^2.
\end{equation}

Now, we can pass to the evaluation of the potential energy given
by the terms related to the AFM exchange interaction. In the
domains with the ferromagnetic order (ferrons), the AFM exchange
leads to the positive contribution to the total energy
\begin{equation}\label{Epot1}
E_{pot1} = 2\left(J_x + J_y\right)S^2n\frac{L_xL_y}{a^2}.
\end{equation}
For the AFM regions, which are free of ferrons, the corresponding
contribution to the energy can be written as
\begin{equation}\label{Epot2}
E_{pot2} = -2\left(J_x + J_y\right)S^2\left(1 -
n\frac{L_xL_y}{a^2}\right).
\end{equation}

Hence, the total potential energy yields
\begin{equation}\label{Epot}
E_{pot} = -2\left(J_x + J_y\right)S^2 + 4\left(J_x + J_y\right)S^2
n\frac{L_xL_y}{a^2}.
\end{equation}

As a result, the total energy related to the formation of FM
droplets has the following form
\begin{eqnarray}\label{Etot}
E_{tot}=-2\left[t_xn + t_yn + \left(J_x + J_y\right)S^2\right]
 + \nonumber \\ n\left[t_x\left(\frac{\pi a}{L_x}\right)^2 +
 t_y\left(\frac{\pi a}{L_y}\right)^2\right]
 + 4\left(J_x + J_y\right)S^2 n\frac{L_xL_y}{a^2}.
\end{eqnarray}

The minimization of energy~\eqref{Etot} with respect to $L_x$ and
$L_y$ gives
\begin{eqnarray}\label{dEtot/dL}
\frac{\partial E_{tot}}{\partial L_x}=-2nt_x\frac{\pi^2
a^2}{L_x^3} + 4\left(J_x + J_y\right)nS^2 \frac{L_y}{a^2} = 0, \nonumber\\
\frac{\partial E_{tot}}{\partial L_y}=-2nt_y\frac{\pi^2
a^2}{L_y^3} + 4\left(J_x + J_y\right)nS^2 \frac{L_x}{a^2} = 0.
\end{eqnarray}

A solution to equations~\eqref{dEtot/dL} reads

\begin{eqnarray}\label{dEtot/dL-sol}
t_x\pi^2 = 2\frac{L_yL_x^3}{a^4}\left(J_x + J_y\right)S^2,\nonumber\\
t_y\pi^2 = 2\frac{L_xL_y^3}{a^4}\left(J_x + J_y\right)S^2.
\end{eqnarray}

Multiplying two equations~\eqref{dEtot/dL-sol} by each other, we
find

\begin{equation}\label{dEtot/dL-sol1}
 \left(\frac{L_yL_x}{a^2}\right)^4 = \frac{\pi^4t_xt_y}{4\left(J_x +
 J_y\right)^2S^4}.
\end{equation}

Now, introducing notation
\begin{equation}\label{t-j_eff}
 t_{eff} = \left(t_xt_y\right)^{1/2}, \quad \bar{J} = \left(J_x +
 J_y\right)S^2,
\end{equation}
we find
\begin{equation*}
\left(\frac{L_yL_x}{a^2}\right)^4 = \Omega^4 =
\frac{\pi^4t_{eff}^2}{4\bar{J}^2}.
\end{equation*}
Thus, the dimensionless volume (area) $\Omega$ of a 2D ferron can
be written as
\begin{equation}\label{2Dvolume}
\Omega=\frac{\pi}{\sqrt{2}}\left(\frac{t_{eff}}{\bar{J}}\right)^{1/2}.
\end{equation}
We get quite a remarkable relationship expressing the volume of a
2D ferron in terms of the $t_{eff}/\bar{J}$ ratio.

Correspondingly, the minimized total energy~\eqref{Etot} takes the
form
\begin{equation}\label{Etot-min}
E_{tot}=-2\left(t_xn + t_yn + \bar{J}\right)
 + 4\pi \sqrt{2} n\left(t_{eff}\bar{J}\right)^{1/2}.
\end{equation}

Introducing the energy of FM polaron by the relationship
\begin{equation}\label{Epol_def}
E_{pol} = E_{tot} + 2\left(t_xn + t_yn + \bar{J}\right),
\end{equation}
we  get finally
\begin{equation}\label{Epol}
E_{pol} = 8n\Omega\bar{J} = 4\pi \sqrt{2}
n\left(t_{eff}\bar{J}\right)^{1/2}.
\end{equation}
It is again worth to notice that the energy of the FM polaron in
2D depends only upon the product of $t_{eff}$ and $\bar{J}$.

\subsection{An elliptical ferron}

Now, we can consider the energy a two-dimensional FM polaron
having the shape of an ellipse. For the same characteristic sizes
(principal axes)of the ferron, its volume in the case of an
ellipse is $\Omega =\pi L_xL_y/a^2$. The corresponding kinetic
energy is again given by Eqs.~\eqref{E_kin}-\eqref{H_kin}. To
solve the Schr\"{o}dinger equation in this geometry, we should
transform an ellipse to a circle. This could be done, for example,
by the dilatation along the $y$ axis: $y =
\tilde{y}\sqrt{t_y/t_x}$. Then, we have

\begin{equation}\label{H_kin_trnsf}
\hat{H}_{kin} = -a^2 t_x\left(\frac{\partial^2}{\partial x^2}
+\frac{\partial^2}{\partial \tilde{y}^2}\right) = -a^2
t_x\Delta_{\tilde{R}},
\end{equation}
where and $\tilde{R}^2 = x^2 + \tilde{y}^2$ and
$\Delta_{\tilde{R}} = \frac{\partial^2}{\partial\tilde{R}^2}
+\frac{1}{\tilde{R}}\frac{\partial}{\partial\tilde{R}}$ is the
radial part of the Laplacian operator in 2D. Thus, the ellipse
$x^2/L_x^2 + y^2/L_y^2 = 1 $ in the 'old' $x, y$ coordinates
transforms to the $x^2 + \tilde{y}^2 = \tilde{R}_{max}^2$ circle
in the 'new' $x, \tilde{y}$ coordinates. From equation for the
circle in terms of 'new' coordinates $x, \tilde{y}$, it is clear
that $\tilde{R}_{max} = L_x$. Hence, we have

\begin{equation}\label{R_tild}
L_x = L_y\sqrt{\frac{t_x}{t_y}} = \tilde{R}_{max},
\end{equation}
and the ferron volume in the initial ('old') coordinates reads
\begin{equation}\label{Omega}
\Omega = \pi\frac{L_xL_y}{a^2} =
\pi\frac{L_x^2}{a^2}\sqrt{\frac{t_y}{t_x}}
\end{equation}

In this case, a solution to the Schr\"{o}dinger
equation~\eqref{SchrEq} has the form $\Psi = J_0(k\tilde{R})$,
where $J_0$ is the Bessel function of zeroth order. The boundary
condition $J_0(k\tilde{R}_{max}) = 0$ yields $k\tilde{R}_{max}  =
j_{0,1} = 2.404 \approx 3\pi/4$, where $j_{0,1}$ is the first zero
of function $J_0$. This means that
\begin{equation}\label{eps0}
\varepsilon_0 = t_x a^2 k^2 =
t_x\left(\frac{j_{0,1}a}{\tilde{R}_{max}}\right)^2 = t_x
\left(\frac{j_{0,1} a}{L_x}\right)^2.
\end{equation}

Then, we have
\begin{equation}\label{Etot_ellip}
E_{tot} = -2\left[t_xn + t_yn + \left(J_x + J_y\right)S^2\right] +
E_{pol},
\end{equation}
where
\begin{equation}\label{Epol_ellip}
E_{pol} = n\left(\frac{j_{0,1} a}{L_x}\right)^2t_x + 4\left(J_x +
J_y\right)S^2 n\frac{\pi L_x^2}{a^2}\sqrt{\frac{t_y}{t_x}}.
\end{equation}
The minimization of polaron energy~\eqref{Epol_ellip} with respect
to $L_x$ gives
\begin{equation}\label{dEpol/dL-ellip}
\frac{\partial E_{pol}}{\partial L_x}=-2nt_x\frac{j_{0,1}^2
a^2}{L_x^3} + 8\left(J_x + J_y\right)S^2 n\frac{\pi
L_x}{a^2}\sqrt{\frac{t_y}{t_x}} = 0.
\end{equation}
Thus, we have (see Eqs.~\eqref{Omega} and \eqref{dEpol/dL-ellip})
\begin{equation}\label{Omega^2}
\Omega^2 = \pi^2\frac{L_x^4}{a^4}\frac{t_y}{t_x} = \frac{\pi
j_{0,1}^2t_x}{4\left(J_x + J_y\right)S^2}\sqrt{\frac{t_y}{t_x}}
\end{equation}

Introducing again $t_{eff}$ and $\bar{J}$ defined by
Eq.~\eqref{t-j_eff}, we find
\begin{equation}\label{Omega-ellips}
\Omega =
\frac{j_{0,1}\sqrt{\pi}}{2}\left(\frac{t_{eff}}{\bar{J}}\right)^{1/2}.
\end{equation}
So, the ferron volume is again expressed as a function of the
universal ratio $t_{eff}/\bar{J}$. Comparing expressions
\eqref{2Dvolume} and \eqref{Omega-ellips} for the volumes of
rectangular and elliptical ferrons, we find
\begin{equation}\label{vol-ratio}
\frac{\Omega_{ellipse}}{\Omega_{rectangle}} =
\frac{j_{0,1}}{\sqrt{2\pi}} \simeq 0.96 < 1.
\end{equation}
This means that the elliptical ferron is a more compact object
(i.e. it has a smaller volume) in comparison to the rectangular
ferron. Accordingly, the energy of elliptical magnetic polaron can
be written in the following form
\begin{equation}\label{Epol_ellip1}
E_{pol} = 8n\Omega\bar{J} = 4nj_{0,1}\sqrt{\pi}
\left(t_{eff}\bar{J}\right)^{1/2}.
\end{equation}
We can see again that the ferron energy depends only upon  the
product of  $t_{eff}$ and $\bar{J}$. Finally, we can compare the
ferron energies for the cases of rectangular and elliptical shapes
using Eqs.~\eqref{Epol} and \eqref{Epol_ellip1}
\begin{equation}\label{energy_ratio_2D}
\frac{E_{pol}^{ellipse}}{E_{pol}^{rectangle}} =
\frac{j_{0,1}}{\sqrt{2\pi}} =
\frac{\Omega_{ellipse}}{\Omega_{rectangle}} \simeq 0.96 < 1
\end{equation}
We see that the ratio of energies turns out to be identical to
the ratio of the volumes. Thus, the elliptical shape of the
ferron is more favorable in energy than the rectangular shape.
First of all, this is caused by a more compact structure of the
elliptical ferron. Another cause was emphasized in
Ref.~\onlinecite{KaK}. The thing is that the elliptical shape of
ferron in 2D has a close similarity  to the one-electron spectrum
characteristic of the empty square lattice: $\varepsilon_p =
p_x^2/2m_x + p_y^2/2m_y$, where $m_x/2 = t_xa^2$ and $m_y/2 =
t_ya^2$. Concluding this section, we can say that the elliptical
shape is the shape most favorable in energy for a FM droplet in
doped anisotropic antiferromagnets with the 2D square lattice.

\section{The shape of FM droplets in the anisotropic 3D case}

Now, we can include the third dimension (which means the inclusion
of $J_z$ and $t_z$) and consider the shape of a FM droplet in a
doped anisotropic antiferromagnet with the 3D cubic lattice. Of
course (having in mind the results of the previous section), we
have to consider FM droplets with the 2D cross-section most
favorable in energy. In other words, we consider the 3D droplets
having the shape of an ellipse in the $x, y$ plane. Then the
problem effectively reduces to the comparison of energies and
volumes of a cylinder and of an ellipsoid of rotation both having
an elliptical cross-section with dimensions $L_x =
L_y\sqrt{\frac{t_x}{t_y}}$ (see Eq.~\eqref{R_tild}).

\subsection{FM droplets of cylindrical shape}

First, let us consider 3D FM droplets of cylindrical shape. The
volume of such a droplet can be written as

\begin{equation}\label{Omega_cyl}
\Omega = \pi\frac{L_xL_yL_z}{a^3} =
\pi\frac{L_x^2}{a^2}\sqrt{\frac{t_y}{t_x}}\frac{L_z}{a}.
\end{equation}

In  this case, the total energy has the form

\begin{eqnarray}\label{Etot_cyl}
E_{tot} = -2\left[t_xn + t_yn + t_zn + \left(J_x + J_y +
J_z\right)S^2\right]  + \nonumber \\ 4\left(J_x + J_y +
J_z\right)S^2\Omega n +t_xn\left(\frac{j_{0,1} a}{L_x}\right)^2 +
t_zn\left(\frac{\pi a}{L_z}\right)^2.
\end{eqnarray}
The polaron energy
\begin{equation}\label{Epol_def_cyl}
E_{pol} = E_{tot}  + 2\left[t_xn + t_yn + t_zn + \left(J_x + J_y +
J_z\right)S^2\right]
\end{equation}
is given by the expression
\begin{eqnarray}\label{Epol_cyl}
E_{pol} = t_xn\left(\frac{j_{0,1} a}{L_x}\right)^2 +
t_zn\left(\frac{\pi a}{L_z}\right)^2 + \nonumber \\
4\left(J_x + J_y + J_z\right)S^2n \pi \frac{L_x^2}{a^2}
\sqrt{\frac{t_y}{t_x}}\frac{L_z}{a}.
\end{eqnarray}

The minimization of polaron energy~\eqref{Epol_cyl} with respect
to $L_x$ and $L_z$ yields
\begin{eqnarray}\label{dEpol/dL_cyl}
\frac{\partial E_{pol}}{\partial L_x} = -2t_x n\frac{j_{0,1}^2
a^2} {L_x^3}+8\bar{J}n\pi\frac{L_x}{a^2}\frac{L_z}{a}
\sqrt{\frac{t_y}{t_x}} = 0,\nonumber \\
 \frac{\partial E_{pol}}{\partial L_z}=-2t_zn\frac{\pi^2a^2}{L_x^3}
+4\bar{J}n\pi\frac{L_x^2}{a^3}\sqrt{\frac{t_y}{t_x}} = 0.
\end{eqnarray}
where we introduced the effective exchange integral for the 3D
case
\begin{equation}\label{Jeff_3D}
 \bar{J} = \left(J_x + J_y + J_z\right)S^2,
\end{equation}
From Eqs.~\eqref{dEpol/dL_cyl}, we get
\begin{eqnarray}\label{dEpol/dL_cyl1}
\frac{j_{0,1}^2 t_x}{4\pi \bar{J}} = \frac{L_x^4L_z}{a^5}
\sqrt{\frac{t_y}{t_x}}, \nonumber \\
\frac{\pi t_z}{2\bar{J}} = \frac{L_x^2L_z^3}{a^5}
\sqrt{\frac{t_y}{t_x}}.
\end{eqnarray}

Squaring the second equation in \eqref{dEpol/dL_cyl1} and dividing
the result by the first equation, we exclude $L_x$ and obtain the
following expression for $L_z$
\begin{equation}\label{Lz_cyl}
L_z = a\left(\frac{\pi^3}
{j_{0,1}^2\bar{J}}\frac{t_z^2}{\sqrt{t_xt_y}}\right)^{1/5}.
\end{equation}
Substituting Eq.~\eqref{Lz_cyl} to the first equation in
\eqref{dEpol/dL_cyl1}, we get
\begin{equation}\label{Lx^4_cyl}
\frac{j_{0,1}^2 t_x}{4\pi \bar{J}} = \frac{L_x^4}{a^4}
\sqrt{\frac{t_y}{t_x}}\left(\frac{\pi^3}
{j_{0,1}^2\bar{J}}\frac{t_z^2}{\sqrt{t_xt_y}}\right)^{1/5}.
\end{equation}
Hence, we have
\begin{equation}\label{Lx_cyl}
L_x = \frac{a}{\sqrt{2}}\left(\frac{j_{0,1}^3}{\pi^2 \bar{J}}
\frac{t_x^2}{\sqrt{t_yt_z}}\right)^{1/5}.
\end{equation}
Using Eqs.~\eqref{Lz_cyl} and \eqref{Lx_cyl}, we find volume
$\Omega_{cyl}$ of the cylindrical ferron
\begin{equation}\label{Omega_cyl}
\Omega_{cyl} = \pi\frac{L_x^2L_z}{a^3}\sqrt{\frac{t_y}{t_x}} =
\frac{(\pi j_{0,1})^{4/5}}{2}
\left(\frac{t_xt_yt_z}{\bar{J}^3}\right)^{1/5}.
\end{equation}
Introducing the effective hopping integral for the 3D case
\begin{equation}\label{t_eff_3D}
 t_{eff} = \left(t_xt_yt_z\right)^{1/3},
\end{equation}
we can rewrite Eq.~\eqref{Omega_cyl} as
\begin{equation}\label{Omega_cyl1}
\Omega_{cyl} = \frac{(\pi j_{0,1})^{4/5}}{2}
\left(\frac{t_{eff}}{\bar{J}}\right)^{3/5}.
\end{equation}

Similar to the the 2D case, we see that the ferron volume in 3D is
also a function of $t_{eff}/\bar{J}$ ratio, where the effective
parameters are given by Eqs.~\eqref{Jeff_3D} and \eqref{t_eff_3D}.

Substituting expressions ~\eqref{Lz_cyl}, ~\eqref{Lx_cyl} , and
~\eqref{Omega_cyl1} for $L_z$, $L_x$, and $\Omega_{cyl}$,
respectively, to the energy of a FM polaron \eqref{Epol_cyl}, we
get
\begin{equation}\label{Epol_cyl_fin}
E_{pol} = 10n\bar{J}\Omega = 5n(\pi j_{0,1})^{4/5}
\left(t_{eff}^3\bar{J}^2\right)^{1/5}
\end{equation}
We see that the polaron energy in the 3D case again depends on the
universal parameters $t_{eff}$ and $\bar{J}$, but the specific
form of this dependence is slightly different:
$\left(t_{eff}^3\bar{J}^2\right)^{1/5}$ in 3D as compared to
$\left(t_{eff}\bar{J}\right)^{1/2}$ in 2D.

\subsection{FM droplets of ellipsoidal shape}

Here, we calculate the volume and the energy of the FM droplet
having the ellipsoidal shape. The volume of the ellipsoidal
droplet in the 3D case is
\begin{equation}\label{Omega_ell}
\Omega_{ell} = \frac{4}{3}\pi\frac{L_xL_yL_z}{a^3}.
\end{equation}

In this case, the total energy of the system has the form
\begin{eqnarray}\label{Etot_cyl}
E_{tot} = -2\left[t_xn + t_yn + t_zn + \left(J_x + J_y +
J_z\right)S^2\right] \nonumber \\+\varepsilon_0n + 4\left(J_x +
J_y + J_z\right)S^2\Omega n.
\end{eqnarray}

Hence the energy of the elliptical FM polaron can be written as
\begin{equation}\label{Epol_ell}
E_{pol} = \varepsilon_0n + 4\bar{J}\Omega n,
\end{equation}
where we again introduce $\bar{J}$ defined by Eq.~\eqref{Jeff_3D}.

As above, energy $\varepsilon_0$ can be found by solving the
corresponding Schr\"{o}dinger equation
\begin{equation}\label{SchrEq_3D}
\hat{H}_{kin}\Psi(x,y,z) = \varepsilon_0\Psi(x,y,z),
\end{equation}
where
\begin{equation}\label{H_kin_3D}
\hat{H}_{kin} = -a^2\left(t_x\frac{\partial^2}{\partial x^2}
+t_y\frac{\partial^2}{\partial y^2} +
t_z\frac{\partial^2}{\partial z^2}\right).
\end{equation}
Using the the dilatation along the $y$ and $z$ axes: $\tilde{y} =
y\sqrt{t_y/t_x}$ and $\tilde{z} = z\sqrt{t_z/t_x}$, we get
\begin{equation}\label{H_kin_3D_trn}
\hat{H}_{kin} = -t_xa^2\Delta_{\tilde{R}}
\end{equation}
in the'new' coordinates $x$, $\tilde{y}$, and $\tilde{z}$. Here,
we have $\tilde{R}^2 = x^2 + \tilde{y}^2 + \tilde{z}^2$ and
$\Delta_{\tilde{R}} = \frac{\partial^2}{\partial\tilde{R}^2}
+2\frac{1}{\tilde{R}}\frac{\partial}{\partial\tilde{R}}$ is the
radial part of the Laplacian operator in 3D.

In these coordinates, a droplet is confined within a sphere of
radius $\tilde{R}_{max} = L_x$. Accordingly, we have
\begin{equation}\label{H_kin_3D}
L_y\sqrt{t_x/t_y} = L_z\sqrt{t_x/t_z} = L_x = \tilde{R}_{max}
\end{equation}
and the droplet volume expressed in terms of initial ('old')
coordinates reads
\begin{equation}\label{Omega_ell_1}
\Omega_{ell} = \frac{4}{3}\pi\frac{L_xL_yL_z}{a^3} =
\frac{4}{3}\pi\left(\frac{L_x}{a}\right)^3
\frac{\left(t_yt_z\right)^{1/2}}{t_x}.
\end{equation}

A solution to the Schr\"{o}dinger equation \eqref{SchrEq_3D} has
the form
\begin{equation}\label{Psi_3D}
\Psi(k\tilde{R}) =\frac{\sin(k\tilde{R})}{(k\tilde{R})}.
\end{equation}

The boundary condition $\Psi(k\tilde{R}_{max}) = 0$ yields
$k\tilde{R}_{max} = \pi$. Hence, we find
\begin{equation}\label{eps_0_ell}
\varepsilon_0 = t_xa^2k^2 = t_xa^2\frac{\pi^2}{\tilde{R}_{max}^2}
= t_xa^2\frac{\pi^2}{L_x^2}
\end{equation}
and the energy of ellipsoidal FM polaron takes the form
\begin{equation}\label{Epol_ell_1}
E_{pol} = t_x\frac{\pi^2a^2}{L_x^2}n
+4\bar{J}n\frac{4}{3}\pi\left(\frac{L_x}{a}\right)^3
\frac{\sqrt{t_yt_z}}{t_x}
\end{equation}
The minimization of polaron energy \eqref{Epol_ell} with respect
to $L_x$ yields
\begin{equation}\label{dEpol_ell/dL}
\frac{\partial E_{pol}}{\partial L_x} =
-2t_x\frac{\pi^2a^2}{L_x^3}n + 16\bar{J}\pi n\left(\frac{L_x}{a}
\right)^2 \frac{1}{a}\frac{\sqrt{t_yt_z}}{t_x} = 0
\end{equation}
As a result, we get the expression for $L_x$
\begin{equation}\label{Lx-ell}
L_x = a\left(\frac{\pi}{8} \frac{t_x^2}
{\bar{J}\sqrt{t_yt_z}}\right)^{1/5}.
\end{equation}
Substituting Eq.~\eqref{Lx-ell} to \eqref{Omega_ell_1}, we find
the volume of an ellipsoidal droplet
\begin{equation}\label{Omega_ell_1}
\Omega_{ell} = \frac{\pi^{8/5}2^{1/5}}{3}
\left(\frac{t_{eff}}{\bar{J}} \right)^{3/5},
\end{equation}
where $t_{eff}$ is again given by Eq.~\eqref{t_eff_3D}. We see
that the volume of the ellipsoidal droplet is also determined by
the dimensionless universal ratio $t_{eff}/\bar{J}$. Dividing
Eq.~\eqref{Omega_ell_1} by Eq.~\eqref{Omega_cyl1}, we obtain the
ratio of volumes for ellipsoidal and cylindrical droplets
\begin{equation}\label{ratio_ell/cyl}
\frac{\Omega_{ell}}{\Omega_{cyl}} =
\frac{4}{3}\left(\frac{\pi}{2j_{0,1}}\right)^{4/5} \simeq 0.95 <
1.
\end{equation}
We see that the ellipsoidal FM droplet is a more compact object
(with a smaller volume) than the cylindrical one.

Substituting expression \eqref{Lx-ell} for $L_x$ to the polaron
energy \eqref{Epol_ell_1}, we find
\begin{equation}\label{Epol_ell_2}
E_{pol} = 10n\bar{J}\Omega_{ell} =
10n\frac{\pi^{8/5}2^{1/5}}{3}\left(t_{eff}^3\bar{J}^2\right)^{1/5}.
\end{equation}
Hence the ratio of energies corresponding to two different shapes
of ferrons is again identically equal to the ratio of their
volumes
\begin{equation}\label{energy_ratio_3D}
\frac{E_{pol}^{ell}}{E_{pol}^{cyl}} =
\frac{\Omega_{ell}}{\Omega_{cyl}} =
\frac{4}{3}\left(\frac{\pi}{2j_{0,1}}\right)^{4/5} \simeq 0.95 <
1.
\end{equation}
Thus, in the 3D case, the ellipsoidal droplet has the lowest
energy  in agreement with the results of Ref. ~\onlinecite{KaK}.

\section{Conclusions and discussion}

We considered the formation and the shape of droplets  in the most
general cases of doped anisotropic 2D and 3D antiferromagnets with
arbitrary values of the electron hopping integrals $t_{\alpha}$
and the AFM exchange integrals $J_{\alpha}$. We found that in the
anisotropic 2D case (when $\alpha = \{x, y\}$ and $t_x \neq t_y,\,
J_x \neq J_y$), the most energetically favorable shape of FM
droplets is an ellipse. In the anisotropic 3D case (when $\alpha =
\{x, y, z\}$ and we include into consideration the third dimension
with $t_z$  and $J_z$), the most energetically favorable shape of
FM droplets is an ellipsoid. Moreover, both the binding energy and
the volume of FM droplets depend in the 2D as well as in the 3D
cases upon only two universal parameters $t_{eff}$ and $\bar{J}$.
In the 2D case, these parameters are  $t_{eff} =
\left(t_xt_y\right)^{1/2}$ and $\bar{J} = \left(J_x +
 J_y\right)S^2$, whereas, in the 3D case, the corresponding
 expressions have the form $t_{eff} =
\left(t_xt_yt_z\right)^{1/3}$ and $\bar{J} = \left(J_x +
 J_y + J_z\right)S^2$.

Note that in the present paper, we considered only the case of
'free' ferrons, which are not strongly localized at donor
impurities. The study of strongly localized ferrons bound to
impurities, especially their shape and the form of the cloud of
magnetic distortions related to them (similar to those described
in Refs.~\onlinecite{Fer1D} and \onlinecite{ogar}) will be the
subject of a  separate publication.

Note also that the situation would be more complicated for FM
droplets in the frustrated triangular or kagome lattices. This is
a case, for example, in an interesting quasi-1D magnetic material
BaCoO$_3$, where the chains of Co$^{4+}$ ions form a triangular
lattice~\cite{BaCoO3,BaCoO3a}.

\section*{Acknowledgments}

The authors acknowledge helpful discussions with D.~Baldomir,
J.~Castro, I.~Gonz\'alez, M.~Hennion, D.I.~Khomskii, S.L.~Ogarkov,
and A.O.~Sboychakov.

The work was supported by the Russian Foundation for Basic
Research, project No.~05-02-17600 and International Science and
Technology Center, grant No.~G1335.

\end{document}